\documentclass[%
 reprint,
 amsmath,amssymb,
]{revtex4-1}

\usepackage{multirow}
\usepackage{graphicx}
\usepackage{dcolumn}
\usepackage{bm}
\usepackage{amsmath}
\usepackage{xcolor}
\usepackage{hyperref}
\usepackage{tikz}
\usetikzlibrary{decorations.pathmorphing}
\usepackage{tikz-3dplot}
\hypersetup{
    citecolor=blue,
    colorlinks=true,
    linkcolor=blue,
    filecolor=blue,      
    urlcolor=blue,
}

\usepackage[english]{babel} 

\begin{document}


\title{Energy-efficient spin injector into semiconductors driven by elastic waves}

\author{Andrei V. Azovtsev$^*$}
\author{Andrei I. Nikitchenko}
\author{Nikolay A. Pertsev}
\affiliation{Ioffe Institute 194021 St. Petersburg Russia\\*azovtsev@mail.ioffe.ru}


\begin{abstract}
Generation of significant spin imbalance in nonmagnetic semiconductors is crucial for the functioning of many spintronic devices, such as magnetic diodes and transistors, spin-based logic gates, and spin-polarized lasers. An attractive design of spin injectors into semiconductors is based on a spin pumping from a precessing ferromagnet, but the classical excitation of magnetization precession by a microwave magnetic field leads to the high power consumption of the device. Here we describe theoretically a spin injector with greatly reduced energy losses, in which the magnetic dynamics is excited by an elastic wave generated in a ferromagnet-semiconductor heterostructure by an attached piezoelectric transducer. To demonstrate the efficient functioning of such an injector, we first perform micromagnetoelastic simulations of the coupled elastic and magnetic dynamics in $\mathrm{Ni}$ films and $\mathrm{Ni}$/$\mathrm{GaAs}$ bilayers traversed by plane longitudinal and shear waves. For thick $\mathrm{Ni}$ films, it is shown that a monochromatic acoustic wave generates a spin wave with the same frequency and wavelength, which propagates together with the driving wave over distances of several micrometers at the excitation frequencies $\nu \approx 10$~GHz close to the frequency of ferromagnetic resonance. The simulations of $\mathrm{Ni}$/$\mathrm{GaAs}$ bilayers with $\mathrm{Ni}$ thicknesses comparable to the wavelength of the injected acoustic wave demonstrate the development of a steady-state magnetization precession at the $\mathrm{Ni}|\mathrm{GaAs}$ interface. The amplitude of such a precession has a maximum at $\mathrm{Ni}$ thickness amounting to three quarters of the wavelength of the elastic wave, which is explained by an analytical model. Using simulation data obtained for the magnetization precession at the $\mathrm{Ni}|\mathrm{GaAs}$ interface, we evaluate the spin current pumped into $\mathrm{GaAs}$ and calculate the spin accumulation in the semiconducting layer by solving the spin diffusion equation. Then the electrical signals resulting from the spin flow and the inverse spin Hall effect are determined via the numerical solution of the Laplace's equation. It is shown that amplitudes of these ac signals near the interface are large enough for experimental measurement, which indicates an efficient acoustically driven spin pumping into $\mathrm{GaAs}$ and rather high spin accumulation in this semiconductor.
\end{abstract}

\maketitle


\setlength{\parindent}{15pt}
\setlength{\parskip}{0pt}

\section{Introduction}\label{sec:intro}
Semiconductors are attractive for the development of spintronic devices due to their large spin diffusion lengths in comparison with transition metals~\cite{Hagele:1998, Kikkawa:1999}, long spin relaxation times~\cite{Bhat:2014}, and the possibility of manipulating the electrons' spin by polarized light~\cite{Putikka:2004, Kokurin:2013}. However, the application of conventional nonmagnetic semiconductors in spintronics requires the generation of an internal spin imbalance by an external stimulus or via an attached magnetic material~\cite{Hirohata:2020}. The simplest method to create such an imbalance would be the direct injection of a spin-polarized charge current from a metallic ferromagnet through Ohmic contact, but the conductance mismatch at the semiconductor-metal interface makes this method inefficient~\cite{Schmidt:2000}. The presence of a thin insulating interlayer acting as a tunnel barrier solves the mismatch problem~\cite{Hanbicki:2002, Jiang:2005, Dash:2009, Kamerbeek:2014}, but requires the fabrication of a high quality interlayer unless the formation of a natural Schottky barrier with the appropriate parameters occurs~\cite{Hanbicki:2002}. Alternatively, the spin imbalance in the semiconductor can be created by bringing it into a direct contact with a precessing ferromagnet~\cite{Brataas:2002, Tserkovnyak:2005}. The resulting spin pumping into the nonmagnetic semiconductor is due to the modulation of the interface scattering matrix by the coherent precession of the magnetization~\cite{Brataas:2002}.

Typically, in spin pumping experiments magnetization dynamics is excited by an external microwave magnetic field with the frequency matching that of the ferromagnetic resonance. Efficient generation of spin currents in normal metals by this technique has been demonstrated experimentally~\cite{Heinrich:2003, Saitoh:2006, Bell:2008, Mosendz:2010, Czeschka:2011, Tashiro:2015}. The spin pumping into semiconductors from metallic ferromagnets~\cite{Ando:2011, Shikoh:2013, Lee:2014, Wang:2017} and ferrimagnetic insulators~\cite{Mendes:2018} subjected to microwave radiation has been revealed as well. However, the power consumption associated with the generation of microwave magnetic fields appears to be rather high, which impedes applications of magnetically driven spin injectors in low-power spintronics. For this reason, alternative spin pumping techniques have been studied during the past decade, one of which is based on the excitation of magnetization dynamics in ferromagnets by injected elastic waves~\cite{Weiler:2011, Uchida:2011, Weiler:2012, Kamra:2015, Polzikova:2016, PRB:2016, APL:2017, Polzikova:2018, PRB:2019, Alekseev:2020}. Since such waves can be generated by a piezoelectric transducer coupled to the ferromagnet and subjected to an ac electric field, the power consumption of elastically driven spin injectors is expected to be comparatively low \cite{Alekseev:2020,Cherepov:2014,Bhaskar:2020}. The experimental and theoretical studies have demonstrated an efficient generation of spin currents in normal metals by surface and bulk acoustic waves, but the strain-driven spin pumping into semiconductors was not investigated so far.

In this paper, we theoretically describe a spin injector into nonmagnetic semiconductors, which employs the spin pumping generated by a dynamically strained ferromagnetic film. The injector has the form of a ferromagnet-semiconductor bilayer coupled to a piezoelectric transducer excited by a microwave voltage. Such a transducer creates a bulk elastic wave propagating across the bilayer, which induces a radio-frequency magnetization precession providing efficient spin pumping into the semiconducting layer. To quantify the elastically driven magnetic dynamics in the ferromagnetic film, we employ the state-of-the-art numerical simulations allowing for the two-way coupling between spins and strains (see Sec.~\ref{sec:method}). The simulations are performed for the $(001)$-oriented $\mathrm{Ni}$ films and $\mathrm{Ni/GaAs}$ bilayers traversed by plane longitudinal and transverse acoustic waves. For thick $\mathrm{Ni}$ films, tightly coupled elastic and magnetic dynamics are described (Sec.~\ref{sec:thick}), which involve the generation of a spin wave carried by the propagating elastic wave. In Sec.~\ref{sec:hetero}, we report the results of numerical simulations performed for $\mathrm{Ni/GaAs}$ bilayers with the $\mathrm{Ni}$ thickness comparable to the wavelength of the propagating elastic wave and discuss the influence of the thickness of the ferromagnetic layer and the excitation frequency on the amplitude of the magnetization precession at the $\mathrm{Ni|GaAs}$ interface (Sec.~\ref{sec:hetero}). Numerical results obtained for the steady-state magnetization precession at the interface are then used to calculate the spin pumping into the $\mathrm{GaAs}$ film and to determine the spin accumulation in the semiconductor by solving the spin diffusion equation (Sec.~\ref{sec:spin_pumping}). It is shown that the proposed injector has a high efficiency ensuring significant spin flux in $\mathrm{GaAs}$, which can be detected experimentally via the inverse spin Hall effect.

\begin{figure}
    \centering

\begin{tikzpicture}[>=stealth,decoration=snake]
\filldraw[fill=orange!20!white] (-3,-2) rectangle (-2,0);
\filldraw[fill=blue!25!white] (-2,-2) rectangle (1,0);
\filldraw[fill=red!15!lightgray] (1,-2) rectangle (4,0);
\filldraw[fill=orange!20!white] (-3,0) -- (-2,1.5) -- (-1,1.5) -- (-2,0) -- cycle;
\filldraw[fill=blue!25!white] (-2,0) -- (-1,1.5) -- (2,1.5) -- (1,0) -- cycle;
\filldraw[fill=red!15!lightgray] (1,0) -- (2, 1.5) -- (5,1.5) -- (4,0) -- cycle;
\filldraw[fill=red!15!lightgray] (4,0) -- (5,1.5) -- (5,-0.5) -- (4,-2) -- cycle;
\draw[thick,->] (-1.5,0.75) -- (-1.5,1.75) node[above] {$y$};
\draw[thick,->] (-1.5,0.75) -- (-2,0) node [above left] {$z$};
\draw[thick,->] (-1.5,0.75) -- (-0.5, 0.75) node [right] {$x$};
\draw[very thick,->,color=blue] (-1.5,0.75) -- (-1.1, 1.7) node [above] {$\textbf{m}$};
\draw[dashed] (-1.1, 1.7) -- (-1.1, 1.7 |- -1.75,0.25);
\draw[dashed] (-1.75,0.25) -- (-1.75,0.25 -| -1.1, 1.7);
\draw[->] (-1.5,0.75) -- (-1.75,0.25 -| -1.1, 1.7);
\draw [thick,domain=-125:-50] plot ({-1.5 + 0.35*cos(\x)}, {0.75 + 0.35*sin(\x)});
\node[above right] at (-1.75,-0.05) {$\psi$};
\node[rotate=90] at (-2.5, -1) {transducer};
\draw[<->] (-2,-1.85) -- (1,-1.85) node[above,midway] {$t_{\text{F}}$};
\draw[<->] (1,-1.85) -- (4,-1.85) node[above,midway] {$t_{\text{N}}$};
\draw[<->] (3.85,0) -- (4.85,1.5) node[left,midway] {$w_{\text{N}}$};
\draw[<->] (3.85,-2) -- (3.85,0) node[left,midway] {$h_{\text{N}}$};
\draw[thick,->,decorate] (-2,-0.9) -- (0.5,-0.9) node[midway,above] {$\textbf{k}$};
\node[below] at (3.25,1.5) {$\mathrm{GaAs}$};
\node[below] at (0.25,1.5) {$\mathrm{Ni}$};

\filldraw[fill=yellow!50!white] (4.3,-0.2) -- (4.9,0.7) -- (4.9,-0.3) -- (4.3,-1.2) -- cycle;
\filldraw[fill=yellow!50!white] (4.3,-0.2) -- (4.2,-0.2) -- (4.2,-1.2) -- (4.3,-1.2) -- cycle;
\filldraw[fill=yellow!50!white] (4.3,-0.2) -- (4.2,-0.2) -- (4.8,0.7) -- (4.9,0.7) -- cycle;
\filldraw[fill=green!25!white] (2, 0.7) -- (2.1, 0.85) -- (2.5, 0.85) -- (2.4, 0.7) -- cycle;
\filldraw[fill=green!25!white] (2, 0.7) -- (2, 0.65) -- (2.4, 0.65) -- (2.4, 0.7) -- cycle;
\filldraw[fill=green!25!white] (2.5, 0.85) -- (2.5, 0.8) -- (2.4, 0.65) -- (2.4, 0.7) -- cycle;
\node[below] at (2.2,0.65) {Fe};
\draw[thick] (2.25, 1) -- (5.5, 1);
\draw[thick] (2.25, 1) -- (2.25, 0.8);
\draw[thick] (5.5, 1) -- (5.5, 0.7);
\filldraw[color=black, fill=white, very thick](5.5, 0.4) circle (0.3);
\node[] at (5.5, 0.4) {$V_{\text{s}}$};
\draw[thick] (5.5, 0.1) -- (5.5, -0.1);
\draw[thick] (5.5, -0.1) -- (4.7, -0.1);
\draw[thick,->,red] (2.1, 0.77) -- (2.4, 0.77);
\end{tikzpicture}

    \caption{\label{fig:setup}
  Ferromagnet-semiconductor heterostructure comprising $\mathrm{Ni}$ and $\mathrm{GaAs}$ layers. An elastic wave (longitudinal or transverse) with the wave vector $\textbf{k}$ is injected into the $\mathrm{Ni}$ layer by the attached piezoelectric transducer. The thicknesses of the $\mathrm{Ni}$ and $\mathrm{GaAs}$ layers are denoted by $t_{\text{F}}$ and $t_{\text{N}}$, respectively. Precessing magnetization in Ni creates a spin imbalance in GaAs, which then produces a measurable voltage $V_{\text{s}}$ between an attached iron probe and a nonmagnetic contact.
  }
\end{figure}
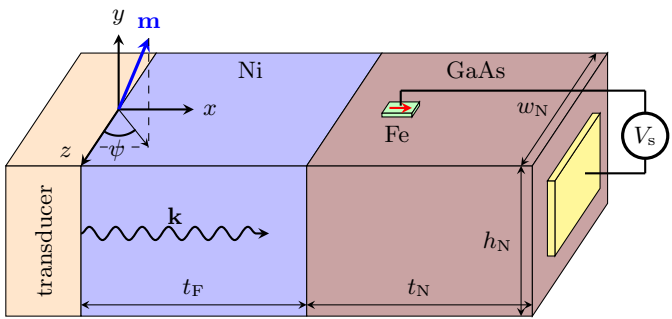

\section{Modeling of magnetoelastic phenomena in ferromagnetic heterostructures}\label{sec:method}

Owing to the magnetoelastic coupling between spins and strains, the excitation of an elastic wave in a ferromagnetic material can induce a precessional motion of the magnetization and the generation of a spin wave \cite{Weiler:2011, Uchida:2011APL, Weiler:2012, Thevenard:2014, Janusonis:2015, Gowtham:2015, Casals:2020}. The backaction of the induced magnetization precession on strain state of the ferromagnet may significantly affect the propagation of the driving elastic wave and lead to the appearance of additional ``secondary'' waves \cite{APL:2017, PRB:2019}. Therefore, the two-way interplay between elastic and magnetic variables \cite{Akhiezer:1958} should be fully taken into account for an accurate modeling of the magnetoelastic phenomena in ferromagnets. Such \textit{micromagnetoelastic modeling} can be realized via the numerical solution of the system of differential equations comprising the elastodynamic equation for the mechanical displacement \textbf{u} and the Landau-Lifshitz-Gilbert (LLG) equation for the magnetization \textbf{M} \cite{Chen:2017, APL:2017, PRB:2019, PRM:2020}. The elastodynamic equation should allow for the magnetoelastic contribution $\delta\sigma_{ij}^{\text{ME}}$ to the mechanical stresses $\sigma_{ij}$ in the ferromagnet, which can be calculated as $\delta\sigma_{ij}^{\text{ME}} = \partial F_{\text{ME}}/\partial \varepsilon_{ij}$, where $F_{\text{ME}}$ is the magnetoelastic energy density, and $\varepsilon_{ij}$ are the elastic strains ($i, j = x, y, z$). The influence of strains on the magnetization orientation can be quantified by adding a magnetoelastic term $\textbf{H}_{\text{ME}} = -(1/\mu_0)\partial F_{\text{ME}}/\partial \textbf{M}$ to the effective magnetic field $\textbf{H}_{\text{eff}}$ involved in the LLG equation ($\mu_0$ being the magnetic constant). For cubic ferromagnets such as nickel, the magnetoelastic contribution to the total energy density F can be written as
\begin{equation}\label{eq:energy_density}
\begin{alignedat}{4}
    F_{\text{ME}} &= B_1 [ (m_x^2-\frac{1}{3})\varepsilon_{xx} + (m_y^2-\frac{1}{3})\varepsilon_{yy} + (m_z^2-\frac{1}{3})\varepsilon_{zz}] \\ &+ 2B_2 \left[ m_x m_y \varepsilon_{xy} + m_x m_z \varepsilon_{xz} + m_y m_z \varepsilon_{yz} \right],
\end{alignedat}
\end{equation}
where $\textbf{m} = \textbf{M}/M_s$ is the unit vector in the magnetization direction, $M_s$ is the saturation magnetization regarded as a strain-independent quantity, and $B_1$, $B_2$ are the magnetoelastic coupling constants \cite{Kittel:1949}.

In this work, we performed micromagnetoelastic simulations of $\mathrm{Ni}$ films and $\mathrm{Ni}$/$\mathrm{GaAs}$ bilayers subjected to a periodic displacement $\textbf{u}(x = 0, t) = \textbf{u}_0(t)$ imposed at the $\mathrm{Ni}$ surface $x = 0$ (Fig. \ref{fig:setup}). Such a displacement models the action of a piezoelectric transducer coupled to $\mathrm{Ni}$ film and generates a plane elastic wave traversing the heterostructure \cite{APL:2017, PRB:2019}. To excite a longitudinal wave characterized by the strain $\varepsilon_{xx}(x, t)$, we introduced the surface displacement with the components $u^0_y = u^0_z = 0$ and $u^0_x = u_{\text{max}}\sin{(2\pi\nu t)}$, while a transverse wave with the shear strain $\varepsilon_{xz}(x, t)$ was created by setting $u^0_x = u^0_y = 0$ and $u^0_z = u_{\text{max}}\sin{(2\pi\nu t)}$. The excitation frequency $\nu$ was varied in a wide range spanning the resonance frequency $\nu_{\text{res}}$ of the coherent magnetization precession in the unstrained $\mathrm{Ni}$ film, which was determined by simulations of the magnetization relaxation to its equilibrium orientation. To ensure the same maximal strain in the elastic wave at any excitation frequency $\nu$, the displacement amplitude $u_{\text{max}}$ was taken to be inversely proportional to $\nu$ \cite{PRB:2019}. Namely, we used the relations $u_{\text{max}} = \varepsilon_{xx}^{\text{max}}/k_L$ and $u_{\text{max}} = 2\varepsilon_{xz}^{\text{max}}/k_T$ for longitudinal and transverse waves, respectively, where $k_L = 2 \pi\nu/c_L$ and $k_T = 2 \pi\nu/c_T$ are the wave numbers of these waves having velocities $c_L$ and $c_T$.

The magnetization dynamics in the $\mathrm{Ni}$ film was quantified using the LLG equation with the effective magnetic field $\textbf{H}_{\text{eff}}$ comprising contributions resulting from the exchange interaction, cubic magnetocrystalline anisotropy, magnetoelastic coupling, Zeeman energy, and dipolar interactions between oscillating spins \cite{PRB:2016}. For numerical calculations, we characterized $\mathrm{Ni}$ by the saturation magnetization $M_s = 4.78\times10^5$ A m$^{-1}$ \cite{Niitsu:2020}, exchange constant $A_{\text{ex}} = 0.85\times10^{-11}$ J m$^{-1}$ \cite{Niitsu:2020}, magnetocrystalline anisotropy constants $K_1 = -5.7\times10^3$ J m$^{-3}$, $K_2 = -2.3\times10^3$ J m$^{-3}$ \cite{Stearns:1986}, magnetoelastic constants $B_1 = 9.2\times10^6$ J m$^{-3}$, $B_2 = 10.7\times10^6$ J m$^{-3}$ \cite{Stearns:1986}, and Gilbert damping parameter of 0.045 \cite{Walowski:2008}. The elastic stiffnesses $c_{11}$, $c_{44}$ and mass densities $\rho$ of $\mathrm{Ni}$ and $\mathrm{GaAs}$, which are involved in their elastodynamic equations of motion, were taken from Ref. \cite{CRC} and listed in Table \ref{tab:tab_constants} together with the velocities $c_L$ and $c_T$ of elastic waves in these materials. No elastic damping was added to the elastodynamic equation in our simulations for the following reasons. First, we are interested in investigating the purely magnetic damping of elastic waves in $\mathrm{Ni}$, and the introduction of the intrinsic elastic damping would obscure simulation results described in Sec. \ref{sec:thick}. Second, in Sec. \ref{sec:hetero} we consider $\mathrm{Ni}$/$\mathrm{GaAs}$ bilayers comprising $\mathrm{Ni}$ films much thinner than the decay lengths of longitudinal and transverse elastic waves in $\mathrm{Ni}$, which are measured to be 5.8 and 29 $\mu$m respectively at the relevant frequency of 9.4 GHz \cite{Homer:1987}.

Micromagnetoelastic simulations were performed with the aid of a homemade software operating with a finite ensemble of nanoscale computational cells. Our software solves the elastodynamic equations of $\mathrm{Ni}$ and $\mathrm{GaAs}$ films by a finite-difference technique with a midpoint derivative approximation and numerically integrates the LLG equation by the projective Runge-Kutta algorithm. We employed a fixed integration step $\delta t = 100$ fs and set the size of cubic computation cells to $2$ nm, which is smaller than the exchange length $l_{\text{ex}} = \sqrt{2A_{\text{ex}}/(\mu_0 M_s^2)} \approx 7.7$ nm of $\mathrm{Ni}$. The system of partial differential equations was appended by appropriate boundary conditions. At the free surface of the $\mathrm{GaAs}$ layer, the stresses $\sigma_{ix}$ were set to zero, and the ``free-surface'' condition $\partial \textbf{m}/\partial x = 0$ was imposed at both boundaries of the $\mathrm{Ni}$ layer. Since a unified ensemble of computational cells covering the whole $\mathrm{Ni}$/$\mathrm{GaAs}$ bilayer was employed in the simulations, the continuity conditions at the $\mathrm{Ni}|\mathrm{GaAs}$ interface were satisfied automatically. The layers comprising the heterostructure were considered infinite in $y-z$ plane and the dynamical quantities were allowed to change only along the $x$ direction.

\begin{table}[]
    \centering
    \begin{tabular}{c||c|c}
         & $\mathrm{Ni}$ & $\mathrm{GaAs}$\\
         \hline
         $c_{11}$ ($10^{11}$ J m$^{-3}$) & 2.481 & 1.188 \\
         $c_{12}$ ($10^{11}$ J m$^{-3}$) & 1.549 & 0.537 \\
         $c_{44}$ ($10^{11}$ J m$^{-3}$) & 1.242 & 0.594 \\
         $\rho$ (kg m$^{-3}$) & 8910 & 5317 \\
         $c_L$ (m s$^{-1}$) & 5277 & 4726 \\
         $c_T$ (m s$^{-1}$) & 3734 & 3344
    \end{tabular}
    \caption{Elastic stiffnesses and mass densities of $\mathrm{Ni}$ and $\mathrm{GaAs}$ \cite{CRC} used in numerical calculations. Velocities $c_L = \sqrt{c_{11}/\rho}$ and $c_T = \sqrt{c_{44}/\rho}$ of longitudinal and transverse elastic waves in these materials are also given for information.}
    \label{tab:tab_constants}
\end{table}

\section{Magnetic dynamics excited by longitudinal and transverse elastic waves in thick N\lowercase{i} films}\label{sec:thick}

An elastic wave perturbs the ferromagnetic state when it creates a nonzero magnetoelastic torque $\textbf{T}_{\text{ME}} = \textbf{M}\times\textbf{H}_{\text{ME}}$ acting on the magnetization $\textbf{M}$. In the case of the longitudinal wave $\varepsilon_{xx}(x, t)$, the effective field $\textbf{H}_{\text{ME}}$ has the only nonzero component $H_x^{\text{ME}} = -2B_1\varepsilon_{xx}m_x/M_s$. Therefore, an external magnetic field $\textbf{H}$ creating the direction cosine $m_x$ in the in-plane magnetized $\mathrm{Ni}$ film should be applied to generate the magnetization dynamics. To additionally stabilize the single-domain state, we introduced the field along the [111] crystallographic direction (easy axis), taking $H_x = H_y = H_z = 1000$ Oe. At such field, the magnetization in the unstrained $\mathrm{Ni}$ film has an elevation angle $\psi \approx 46^{\circ}$ ($m_x = 0.137$) and equal projections on [100] and [010] directions ($m_y = m_z = 0.70044$). The same magnetic field was used in the simulations of magnetic dynamics induced by the shear acoustic wave $\varepsilon_{xz}(x, t)$, which facilitates the comparison of results obtained for two types of elastic perturbations. It should be noted that shear waves impose nonzero magnetoelastic torque $\textbf{T}_{\text{ME}}$ even on the in-plane magnetized ferromagnetic films. Indeed, the effective field $\textbf{H}_{\text{ME}}$ created by the strain $\varepsilon_{xz}$ has two components, $H_z^{\text{ME}} = -2B_2\varepsilon_{xz}m_x/M_s$ and $H_x^{\text{ME}} = -2B_2\varepsilon_{xz}m_z/M_s$, and the latter differs from zero even at $m_x = 0$ when $m_z \ne 0$. However, simulations show that the amplitude of the magnetization precession increases significantly when both $H_x^{\text{ME}}$ and $H_z^{\text{ME}}$ differ from zero.

At the chosen magnetic field $\textbf{H}$, the resonance frequency $\nu_{\text{res}}$ of the unstrained $\mathrm{Ni}$ film with in-plane dimensions much larger than the film thickness $t_{\text{F}}$ was found to be $9.6$ GHz. Accordingly, the excitation frequency was varied in a wide range around 10 GHz. By selecting the appropriate amplitudes $u_{\text{max}}(\nu)$ of the surface displacements $u_x^0(t)$ or $u_z^0(t)$, we created the maximal strains $\varepsilon^{\text{max}}_{xx} = \varepsilon^{\text{max}}_{xz} = 10^{-4}$ in the excited elastic waves near the $\mathrm{Ni}$ surface $x=0$. Simulations were first performed for $\mathrm{Ni}$ films with thicknesses $t_{\text{F}}$ much larger than the wave lengths $\lambda_L = c_L/\nu$ and $\lambda_T = c_T/\nu$ of the pure elastic waves, which amount to $\lambda_L = 550$ nm and $\lambda_T = 389$ nm at the resonance excitation $\nu = \nu_{\text{res}}$. This allows the observation of several wave periods inside the ferromagnetic film. Since in this section we concentrate on the propagation of the elastic waves in $\mathrm{Ni}$ and their interaction with the magnetic subsystem, the simulation time was limited by the period needed for the wave to reach the opposite boundary of the $\mathrm{Ni}$ film. The effects of the wave reflection from the $\mathrm{Ni}|\mathrm{GaAs}$ interface are discussed in Section \ref{sec:hetero}.

\begin{figure}
  \begin{minipage}{1.0\linewidth}
    \center{
    \includegraphics[width=0.8\linewidth]{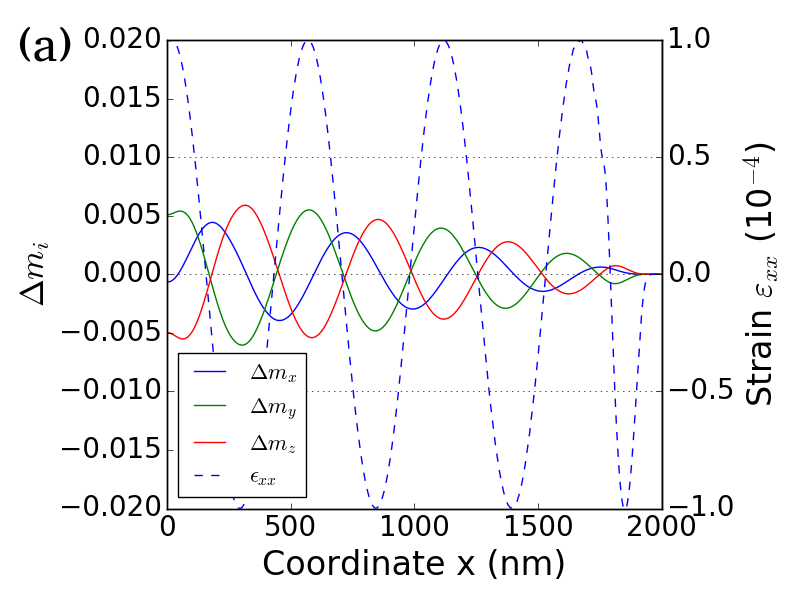}
    }
  \end{minipage}
  \begin{minipage}{1.0\linewidth}
    \center{
    \includegraphics[width=0.8\linewidth]{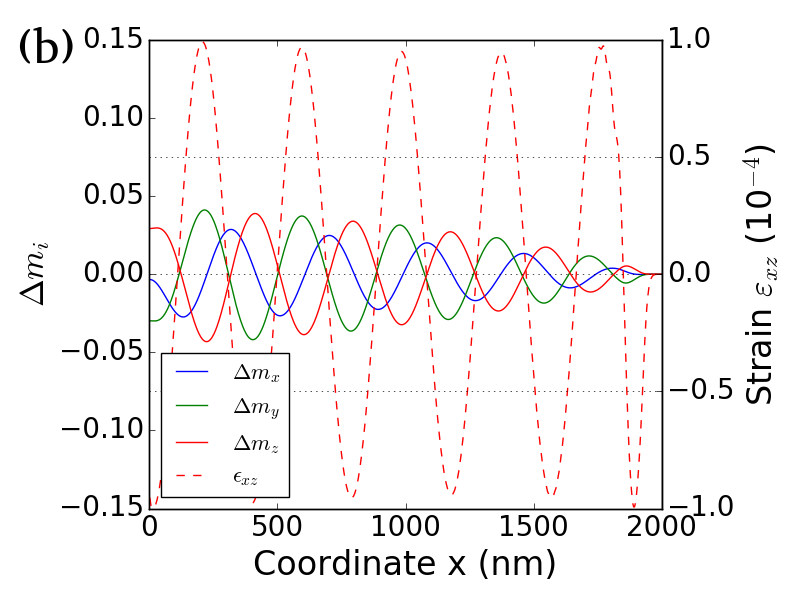}
    }
  \end{minipage}
    \caption{\label{fig:waves_all}
  Spatial distributions of strains in the driving longitudinal (a) and transverse (b) elastic waves and strain-induced variations of the magnetization direction cosines $m_i$ in the 2-$\mu$m-thick $\mathrm{Ni}$ film. Excitation frequency $\nu$ is equal to the resonance frequency $\nu_{\text{res}} = 9.6$ GHz, and snapshots are taken at 0.37 ns (a) and 0.52 ns (b).}
\end{figure}

\begin{figure}[t]
  \begin{minipage}{1.0\linewidth}
    \center{
    \includegraphics[width=0.8\linewidth]{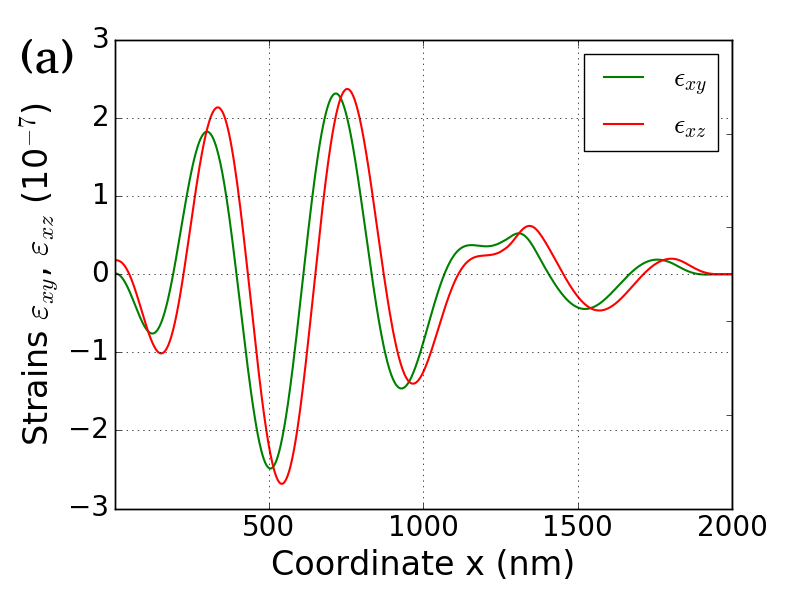}
    }
  \end{minipage}
  \begin{minipage}{1.0\linewidth}
    \center{
    \includegraphics[width=0.8\linewidth]{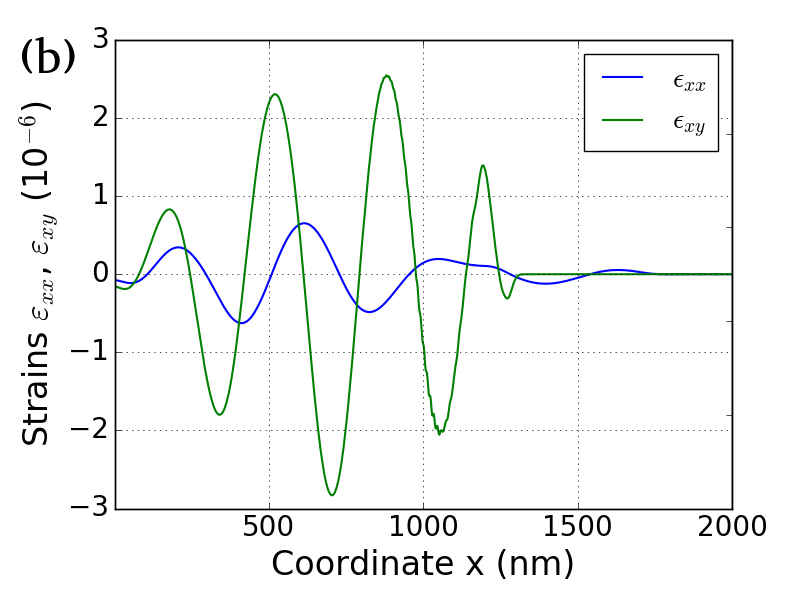}
    }
  \end{minipage}
    \caption{\label{fig:waves_secondary}
  Secondary elastic waves generated by the magnetization precession induced by the primary longitudinal (a) and transverse (b) waves in the 2-$\mu$m-thick $\mathrm{Ni}$ film. Snapshots are taken at 0.37 ns (a) and 0.35 ns (b).}
\end{figure}

The simulations of the coupled elastic and magnetic dynamics in thick $\mathrm{Ni}$ films confirmed the creation of periodic, almost sinusoidal elastic waves at all studied excitation frequencies $\nu$ (see Fig. \ref{fig:waves_all}). The wave emerges at the $\mathrm{Ni}$ surface and propagates away with the velocity $c_L$ or $c_T$ characteristic of a pure elastic wave despite an inhomogeneous magnetization precession excited by the magnetoelastic torque $\textbf{T}_{\text{ME}}$ (see Fig. \ref{fig:waves_all}). However, the strain-induced precession manifests itself in the generation of the additional elastic waves caused by the magnetoelastic feedback (see Fig. \ref{fig:waves_secondary}). These secondary strain waves were already revealed by the micromagnetoelastic simulations performed for Fe$_{81}$Ga$_{19}$ and CoFe$_2$O$_4$ films \cite{APL:2017,PRB:2019}, but were not detected in the simulations of the propagation of longitudinal elastic waves in $\mathrm{Ni}$ \cite{Chen:2017}. When the driving wave is a longitudinal one, two transverse secondary waves with the strains $\varepsilon_{xy}(x, t)$ and $\varepsilon_{xz}(x, t)$ having amplitudes $\sim10^{-7}$ appear in $\mathrm{Ni}$. Their profiles depend on the position in the film, exhibiting a peculiar behavior similar to that of the secondary waves arising in CoFe$_2$O$_4$ excited by longitudinal elastic waves \cite{PRB:2019}. This behavior is caused by the interference of the two components of each secondary wave, which have the form of a shear wave with the wavelength $\lambda_T$ and velocity $c_T$ freely propagating from the $\mathrm{Ni}$ surface and a forced shear wave with the wavelength $\lambda_L$ and velocity $c_L$ generated in the whole driving longitudinal wave. When the driving wave is a transverse one [$\varepsilon_{xz}(x, t)$], a longitudinal secondary wave $\varepsilon_{xx}(x, t)$ and another shear wave $\varepsilon_{xy}(x, t)$ with amplitudes $\sim10^{-6}$ are generated by the magnetization precession. Similarly to the aforementioned situation, the longitudinal wave appears to be a superposition of a free wave with the wavelength $\lambda_L$ and velocity $c_L$ and a forced wave with the parameters $\lambda_T$ and $c_T$. In contrast, the secondary shear wave can be regarded as a single wave because its wavelength $\lambda_T$ and velocity $c_T$ match those of the driving wave.

The magnetization precession induced by the primary elastic wave also affects its propagation at long distances from the $\mathrm{Ni}$ surface. A careful evaluation of the local strain amplitudes $\varepsilon_{xx}^{\text{max}}$ and $\varepsilon_{xz}^{\text{max}}$ in the driving longitudinal and transverse waves reveals that they decrease with the increasing distance $x$ from the $\mathrm{Ni}$ surface. This decay is caused by the energy transfer to the magnetic subsystem, where the strain-driven magnetization precession is hindered by the Gilbert damping \cite{PRB:2019}. The analysis of the simulation results shows that the dependences $\varepsilon_{xx}^{\text{max}}(x)$ and $\varepsilon_{xz}^{\text{max}}(x)$ can be fitted by an exponential function $e^{-x/L_{\text{dec}}}$, where the decay length $L_{\text{dec}}$ depends on the wave frequency $\nu$. At the resonance excitation $\nu = 9.6$ GHz, $L_{\text{dec}}$ amounts to approximately 350 $\mu$m for the longitudinal wave and about 19 $\mu$m for the shear wave. This finding explains why no damping of magnetic origin was detected in simulations of the propagation of longitudinal elastic waves in $\mathrm{Ni}$ through a short distance of 300 nm \cite{Chen:2017}. At the same time, it was shown experimentally that surface acoustic waves (SAWs) can propagate in a $\mathrm{Ni}$ film over the distance of several millimeters \cite{Casals:2020}. This absence of significant damping observed for the studied SAWs with frequencies not exceeding 500 MHz is very different from the results of Homer et al. \cite{Homer:1987}, who reported the decay length $L_{\text{dec}} \approx 5.8~\mu$m for the longitudinal wave with the frequency of 9.4 GHz in $\mathrm{Ni}$. The reason for such a difference most probably lies in a drastic reduction of damping, which should happen when the frequency of the elastic wave changes from about 10 GHz to several hundreds of MHz. As for the damping of transverse elastic waves in $\mathrm{Ni}$, our results show that the magnetic damping of elastic waves ($L_{\text{dec}} \approx 19~\mu$m) could be stronger than the damping of electronic origin (measured $L_{\text{dec}} \approx 29~\mu$m \cite{Homer:1987}) at wave frequencies around 10 GHz.

Most important of all, the magnetoelastic interaction leads to the formation of a spin wave tightly coupled to the driving elastic wave. The spin waves predicted by our simulations have sinusoidal time dependences under both types of elastic excitation, which differs from a non-sinusoidal time dependence reported in Ref. \cite{Chen:2017} for the spin wave generated by the longitudinal acoustic wave having near-resonance frequency of 10 GHz. Regarding the spin wave amplitude, the transverse elastic wave appears to be much more efficient for the generation of spin waves in $\mathrm{Ni}$ than the longitudinal wave at the chosen equilibrium magnetization orientation [compare panels (a) and (b) in Fig. \ref{fig:waves_all}]. Similarly to elastically generated spin waves in Fe$_{81}$Ga$_{19}$ and CoFe$_2$O$_4$ films \cite{APL:2017,PRB:2019}, the spin wave propagating in a thick $\mathrm{Ni}$ film has the same frequency and wavelength as the driving strain wave. Since both waves (spin and elastic) travel with the same velocity $c_L$ or $c_T$ and obey a purely elastic dispersion relation $k_L = 2\pi\nu/c_L$ or $k_T = 2\pi\nu/c_T$, the driving wave acts as a carrier of the spin wave having a forced character. Furthermore, the decay length of the spin wave carried by the longitudinal acoustic wave matches the decay length $L_{\text{dec}} \approx 350~\mu$m of the latter in our simulations, which agrees with the behavior predicted for CoFe$_2$O$_4$ films \cite{PRB:2019}. However, in the case of the excitation by a shear wave, the spin-wave decay length is significantly smaller than that of the driving wave, being roughly 9 $\mu$m instead of 19 $\mu$m. Despite this peculiarity, the acoustically driven spin waves with frequencies $\nu \approx 10$ GHz can still propagate in Ni over long distances of several micrometers, which is important for magnon spintronics.

\section{Magnetoelastic dynamics in N\lowercase{i}/G\lowercase{a}A\lowercase{s} bilayers}\label{sec:hetero}

Now we turn our attention to $\mathrm{Ni}$/$\mathrm{GaAs}$ bilayers
\begin{figure}[t]
  \begin{minipage}{1.0\linewidth}
    \center{
    \includegraphics[width=0.7\linewidth]{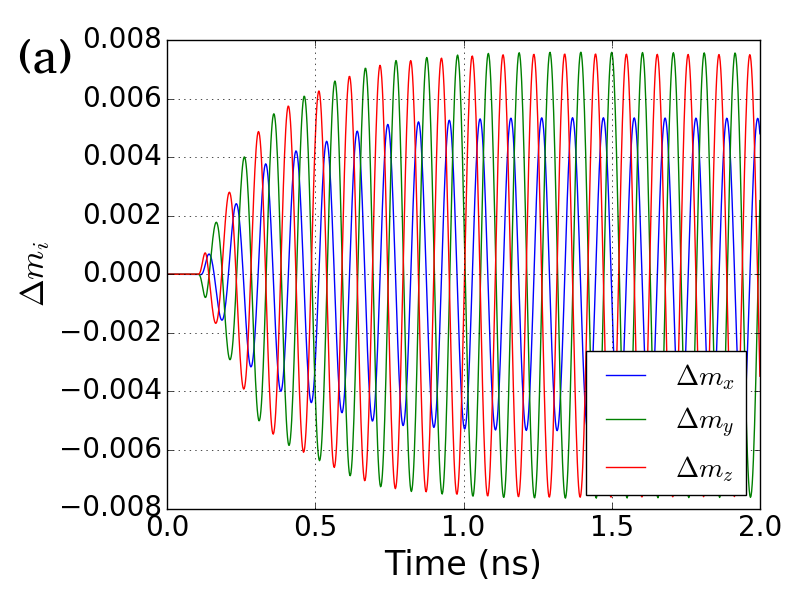}
    }
  \end{minipage}
  \begin{minipage}{1.0\linewidth}
    \center{
    \includegraphics[width=0.7\linewidth]{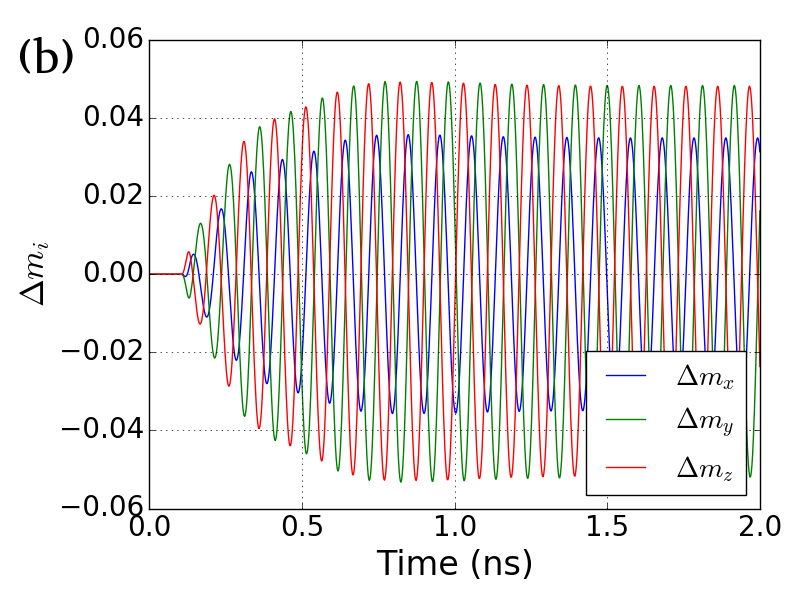}
    }
  \end{minipage}
    \caption{\label{fig:precession_interface}
  Time dependence of the magnetization precession at the $\mathrm{Ni}|\mathrm{GaAs}$ interface excited by longitudinal (a) or transverse (b) elastic waves with the frequency $\nu = \nu_{\text{res}} = 9.6$ GHz. Thickness $t_{\text{F}}$ of $\mathrm{Ni}$ film in each case is equal to a corresponding wavelength $\lambda_L$ or $\lambda_T$.}
\end{figure}
comprising relatively thin $\mathrm{Ni}$ layers with the thickness $t_{\text{F}}$ comparable to the wavelength of the driving elastic wave with the frequency $\nu \sim \nu_{\text{res}}$, which are most suitable for applications in miniature spin injectors (see Sec. \ref{sec:spin_pumping}). The magnetoelastic dynamics in such bilayers is of a more complicated character due to reflections of the elastic waves from the $\mathrm{Ni}|\mathrm{GaAs}$ interface and the $\mathrm{GaAs}$ free surface. Fortunately, owing to similar acoustic impedances of $\mathrm{Ni}$ and $\mathrm{GaAs}$, the transmittance of the driving longitudinal or transverse wave through the $\mathrm{Ni}|\mathrm{GaAs}$ interface is close to unity (about 0.9 with respect to energy). In contrast, the driving wave fully reflects from the $\mathrm{GaAs}$ free surface, and the reflected wave strongly disturbs the magnetization dynamics when it penetrates back into the $\mathrm{Ni}$ layer. In order to avoid this complication, we imparted a strong artificial elastic damping to $\mathrm{GaAs}$, which is sufficient to force any elastic wave to vanish before it reaches the free surface, but does not change significantly the strain dynamics at the $\mathrm{Ni}|\mathrm{GaAs}$ interface.

\begin{figure}
    \center{\includegraphics[width=1\linewidth]{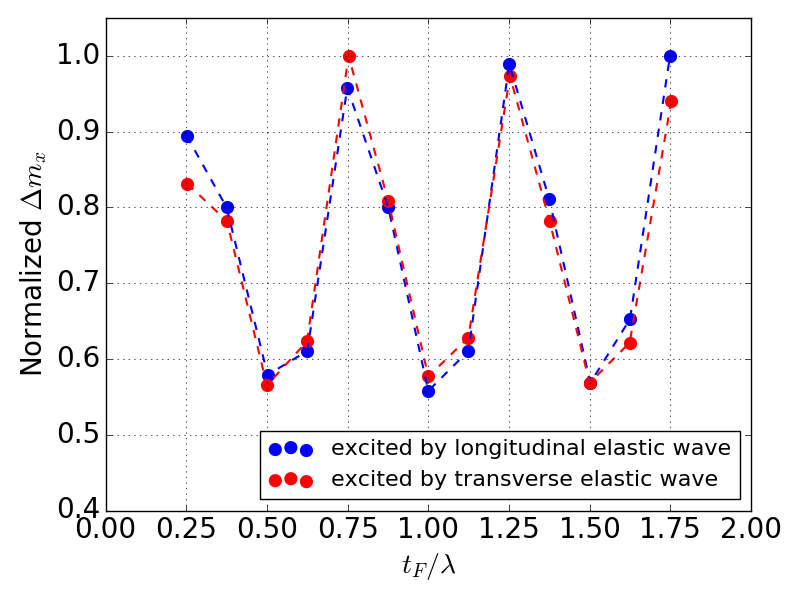}}
    \caption{\label{fig:thickness_precession}
  Amplitude of the magnetization precession at the $\mathrm{Ni}|\mathrm{GaAs}$ interface as a function of the $\mathrm{Ni}$ thickness $t_{\text{F}}$ normalized by the wavelength $\lambda$ of the driving longitudinal or transverse elastic wave. The plots show the maximal change $\Delta m_x$ of the out-of-plane direction cosine $m_x(x = t_{\text{F}}, t)$ normalized by the largest value of $\Delta m_x$ in the studied thickness range. The excitation frequency equals $\nu_{\text{res}} = 9.6$ GHz.}
\end{figure}

The simulations demonstrated that the elastically driven magnetization dynamics in $\mathrm{Ni}$ layers with the thicknesses $t_{\text{F}}$ about $\lambda_L$ or $\lambda_T$ remains to be highly inhomogeneous at the resonance excitation $\nu = \nu_{\text{res}}$. Initially the magnetic dynamics has the form of a spin wave, but it assumes a complex character after several reflections of the driving elastic wave from the boundaries of the $\mathrm{Ni}$ layer. However, near the interface the magnetization precesses with a constant frequency and amplitude in a steady-state regime, which settles in after a transition period of about 1 ns (Fig. \ref{fig:precession_interface}). Performing a series of simulations at different thicknesses of $\mathrm{Ni}$ layers, we found that the amplitude of the magnetization precession at the interface has local maxima at Ni thicknesses amounting to 0.25, 0.75, 1.25 and 1.75 of the wavelength $\lambda_L$ or $\lambda_T$ (Fig. \ref{fig:thickness_precession}). This result differs from that obtained for Fe$_{81}$Ga$_{19}$/Au and CoFe$_2$O$_4$/Pt bilayers in our previous works \cite{APL:2017, PRB:2019}, where such amplitude maximizes at the ferromagnet’s thickness equal to one wavelength of the driving elastic wave.

To understand the revealed behavior of ferromagnetic-nonmagnetic (F/N) bilayers, we investigated the dependence of the strain amplitude at the interface on the thickness $t_{\text{F}}$ of the F layer. The analysis of the results of simulations showed that in all studied bilayers the precession amplitude in the steady-state regime becomes maximal whenever the strain amplitude maximizes. Therefore, we considered a general elasticity problem of finding the strain distribution in an elastic F/N bilayer subjected to a periodic surface displacement $u_i^{\text{F}}(x=0, t) = u_{\text{max}} e^{-i\omega t}$. Despite multiple reflections of the elastic waves from the boundaries of the F layer, the steady-state solution for the elastic displacement $u_i^{\text{F}}(x, t)$ inside this layer can be written as a superposition of two waves with the same frequency $\omega = 2\pi\nu$. Indeed, due to the principle of superposition in linear elasticity any number of interfering sinusoidal waves with the same frequency but different amplitudes and phases produce another sinusoidal wave of the same frequency with its own amplitude and phase \cite{Sadd}. Hence, we can write 
\begin{equation}\label{eq:wave_F}
u_i^{\text{F}}(x,t) = A_i^{\text{F}}e^{i(k_i^{\text{F}}x-\omega t)} + B_i^{\text{F}}e^{-i(k_i^{\text{F}}x+\omega t)},
\end{equation}
where the first term corresponds to the waves propagating towards the F$|$N interface, while the second term describes the waves reflected from the F$|$N interface; $A_i^{\text{F}}$ and $B_i^{\text{F}}$ are the unknown amplitudes of these waves, and $k_i^{\text{F}}$ is the wavenumber of the longitudinal ($i = x$) or transverse ($i = y$ or $z$) wave in the F layer. 
\begin{figure}
    \center{\includegraphics[width=1\linewidth]{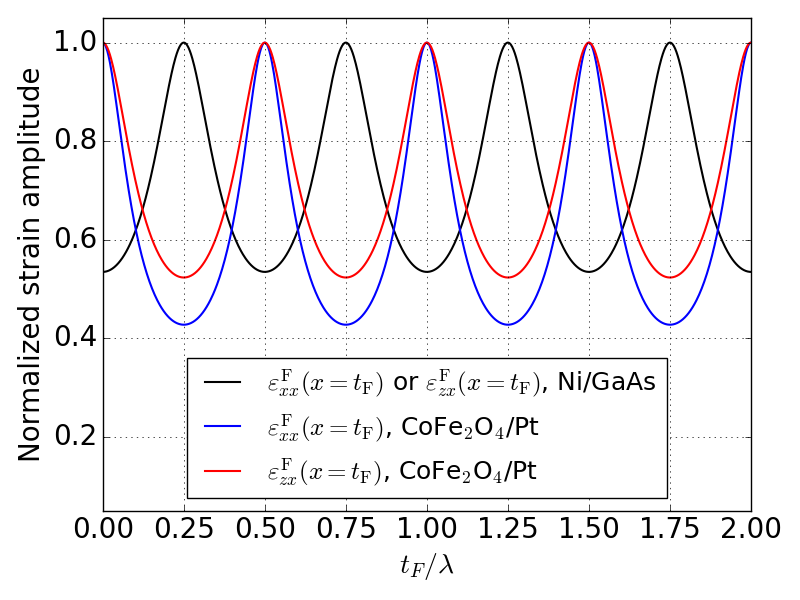}}
    \caption{\label{fig:thickness_strain}
  Expected thickness dependences of the strain amplitudes at the interface between $\mathrm{Ni}$ and $\mathrm{GaAs}$ layers and CoFe$_2$O$_4$ and Pt layers. The amplitude of each strain is normalized by its maximal value. The thickness $t_{\text{F}}$ of the ferromagnetic layer is normalized by the wavelength $\lambda$ of the excited longitudinal or transverse elastic wave. The excitation frequency equals 9.6 GHz ($\mathrm{Ni}$/$\mathrm{GaAs}$) and 11 GHz (CoFe$_2$O$_4$/Pt).
  }
\end{figure}
Since we neglect the reflections from the free surface of the N layer, only the transmitted elastic wave exists in it, and the displacement $u_i^{\text{N}}(x,t)$ has the form
\begin{equation}\label{eq:wave_N}
u_i^{\text{N}}(x,t) = A_i^{\text{N}}e^{i(k_i^{\text{N}}x-\omega t)},
\end{equation}
where $A_i^{\text{N}}$ and $k_i^{\text{N}}$ are the amplitude and wavenumber of the transmitted wave. The mechanical boundary conditions at the F$|$N interface $x = t_{\text{F}}$ yield the displacement continuity $u_i^{\text{F}}(x=t_{\text{F}},t) = u_i^{\text{N}}(x=t_{\text{F}},t)$ and the stress continuity $\sigma_{ix}^{\text{F}}(x=t_{\text{F}},t) = \sigma_{ix}^{\text{N}}(x=t_{\text{F}},t)$. In our model case, the stresses are given by the relations $\sigma_{ix}^{\text{F}}(x,t) = c_{\alpha\alpha}^{\text{F}}(1/\sqrt{\alpha})\partial/\partial x [u_i^{\text{F}}(x,t)]$ and $\sigma_{ix}^{\text{N}}(x,t) = c_{\alpha\alpha}^{\text{N}}(1/\sqrt{\alpha})\partial/\partial x [u_i^{\text{N}}(x,t)]$, where $c_{\alpha\alpha}^{\text{F}}$ and $c_{\alpha\alpha}^{\text{N}}$ are the elastic stiffnesses of the F and N layers, respectively ($\alpha = 1$ at $i = x$ and $\alpha = 4$ at $i = y$ or $z$). Combining the boundary conditions at the F$|$N interface and the F surface $x = 0$ and using Eqs. (\ref{eq:wave_F}) and (\ref{eq:wave_N}), one can derive analytic relations for the unknown amplitudes $A_i^{\text{F}}$, $B_i^{\text{F}}$, and $A_i^{\text{N}}$. The substitution of these relations back to Eqs. (\ref{eq:wave_F}) and (\ref{eq:wave_N}) yields the formulae for the displacements $u_i^{\text{F}}(x,t)$ and $u_i^{\text{N}}(x,t)$, which render possible to calculate the strains $\varepsilon_{ix}^{\text{F}} = (1/\sqrt{\alpha})\partial u_i^{\text{F}}/\partial x$ and $\varepsilon_{ix}^{\text{N}} = (1/\sqrt{\alpha})\partial u_i^{\text{N}}/\partial x$ in the F and N layers. For the strains $\varepsilon_{ix}^{\text{F}}(x=t_{\text{F}},t)$ at the F$|$N interface after some mathematical manipulations we obtain
\begin{equation}\label{eq:strain_at_interface}
    \varepsilon_{ix}^{\text{F}}(x = t_{\text{F}}, t) =  \frac{2i\varepsilon_{ix}^{\text{max}}Z_{\alpha}e^{-i\omega t}}{(1+Z_{\alpha})F^- + (1-Z_{\alpha})F^+},
\end{equation}
where
\begin{equation*}
    Z_{\alpha} = \sqrt{\frac{   c_{\alpha\alpha}^{\text{N}}\rho_{\text{N}}  }{c_{\alpha\alpha}^{\text{F}}\rho_{\text{F}}}}, F^+ = e^{i k_i^{\text{F}} t_{\text{F}} }, F^- = e^{-i k_i^{\text{F}} t_{\text{F}} }. 
\end{equation*}

Equation (\ref{eq:strain_at_interface}) shows that the amplitude of $\varepsilon_{ix}^{\text{F}}(x=t_{\text{F}},t)$ depends on the input strain $\varepsilon_{ix}^{\text{max}} = (1/\sqrt{\alpha})u_{\text{max}}k_i^{\text{F}}$, the relative thickness $t_{\text{F}}/\lambda_L$ or $t_{\text{F}}/\lambda_T$ of the F layer, and the dimensionless parameter $Z_{\alpha}$ of the F/N bilayer, which is governed by the elastic stiffnesses and densities of the involved materials. Using Eq. (\ref{eq:strain_at_interface}), we calculated the dependences of the discussed strain amplitudes on the relative thickness of the F layer for $\mathrm{Ni}$/$\mathrm{GaAs}$ and CoFe$_2$O$_4$/Pt bilayers subjected to the resonance excitation $\nu = \nu_{\text{res}}$. The results presented in Fig. \ref{fig:thickness_strain} show that, for a given bilayer, the amplitudes of $\varepsilon_{xx}^{\text{F}}(x=t_{\text{F}},t)$ and $\varepsilon_{zx}^{\text{F}}(x=t_{\text{F}},t)$ normalized by their maximal values follow similar curves (almost identical in case of $\mathrm{Ni}$/$\mathrm{GaAs}$) when plotted as a function of $t_{\text{F}}/\lambda_L$ and $t_{\text{F}}/\lambda_T$, respectively. However, the maximal strain amplitude is reached at the thicknesses $t_{\text{F}} = (0.25 + 0.5n)\lambda$ in $\mathrm{Ni}$/$\mathrm{GaAs}$ bilayers and at $t_{\text{F}} = (0.5 + 0.5n)\lambda$ in CoFe$_2$O$_4$/Pt ones ($\lambda = \lambda_L$ or $\lambda_T$, $n = 0, 1, 2, 3$...). These conditions explain dissimilar results of our micromagnetoelastic simulations performed for $\mathrm{Ni}$/$\mathrm{GaAs}$ and CoFe$_2$O$_4$/Pt bilayers, which showed that the precession amplitude at the interface has a maximum at $t_{\text{F}} = 0.75\lambda$ ($\mathrm{Ni}$/$\mathrm{GaAs}$) and $t_{\text{F}} = \lambda$ (CoFe$_2$O$_4$/Pt). Furthermore, the analysis of Eq. (\ref{eq:strain_at_interface}) reveals that the character of the strain-amplitude thickness dependence is governed by the magnitude of the parameter $Z_{\alpha}$. Namely, when $Z_{\alpha} < 1$, the strain amplitude maximizes at $t_{\text{F}} = (0.25 + 0.5n)\lambda$ as it happens in the $\mathrm{Ni}$/$\mathrm{GaAs}$ bilayers ($Z_1 \approx Z_4 \approx 0.53$), whereas at $Z_{\alpha} > 1$ the optimal thicknesses satisfy the condition $t_{\text{F}} = (0.5 + 0.5n)\lambda$ holding for the CoFe$_2$O$_4$/Pt bilayers ($Z_1 \approx 2.34$, $Z_4 \approx 1.91$). The derived simple criteria open the possibility to predict the optimal thickness of the ferromagnetic layer that maximizes the strain and precession amplitudes at the interface for any F/N bilayer.

\begin{figure}
    \center{\includegraphics[width=1\linewidth]{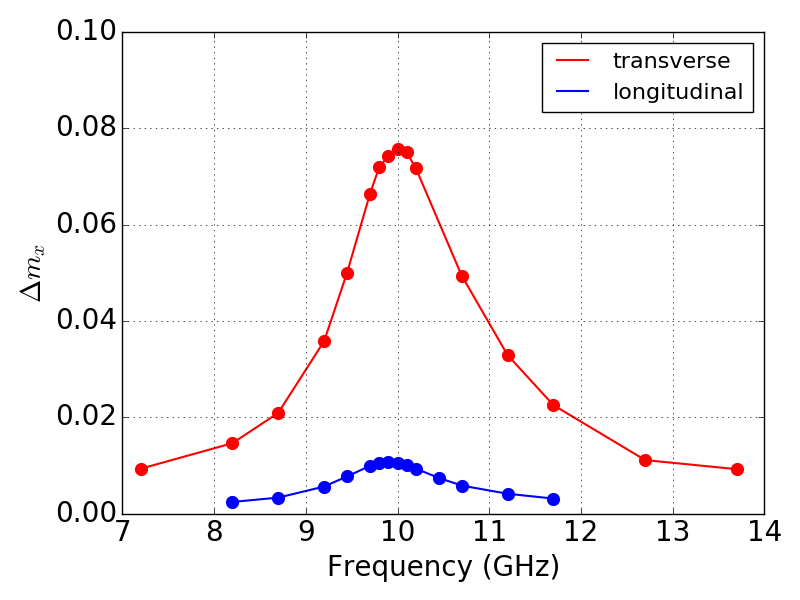}}
    \caption{\label{fig:frequency}
  Dependence of the amplitude of the magnetization precession at the $\mathrm{Ni}|\mathrm{GaAs}$ interface on the excitation frequency $\nu$. The points show the maximal deviation $\Delta m_x(\nu)$ of the out-of-plane magnetization direction cosine $m_x$ from its equilibrium value. The thickness of the $\mathrm{Ni}$ layer equals three quarters of the wavelength of the driving longitudinal or transverse elastic wave.
  }
\end{figure}

In conclusion of this section, we discuss the dependence of the amplitude of the magnetization precession at the $\mathrm{Ni}|\mathrm{GaAs}$ interface on the excitation frequency $\nu$. For the optimal $\mathrm{Ni}$ thickness $t_{\text{F}} = 0.75 \lambda$ and the driving waves with the initial strain amplitudes $\varepsilon_{xx}^{\text{max}} = \varepsilon_{xz}^{\text{max}} = 10^{-4}$, the simulations predict that the maximal deviation $\Delta m_x(\nu)$ of the magnetization direction cosine $m_x$ from the equilibrium value varies with the frequency as shown in Fig. \ref{fig:frequency}. It can be seen that $\Delta m_x(\nu)$ reaches a peak at a frequency $\nu_{\text{max}}$ slightly higher than the resonance frequency $\nu_{\text{res}} = 9.6$ GHz. Namely, $\nu_{\text{max}}$ amounts to 9.9 GHz for the precession excited by the longitudinal elastic waves and to 10 GHz for that induced by the transverse waves. In agreement with the results described in Sec. \ref{sec:thick}, the shear waves appear to be much more efficient for the excitation of the magnetization precession at the $\mathrm{Ni}|\mathrm{GaAs}$ interface (see Fig. \ref{fig:frequency}).

\section{Spin pumping into G\lowercase{a}A\lowercase{s} layer}\label{sec:spin_pumping}
The magnetization precession occurring near the interface between the ferromagnet and a nonmagnetic conductor generates the spin pumping into the latter~\cite{Tserkovnyak:2005}. Using the results obtained for the magnetization dynamics $\mathbf{m}(x = t_\mathrm{F}, t)$ induced by the elastic waves at the $\mathrm{Ni}|\mathrm{GaAs}$ interface, we can calculate the spin current flowing in the $\mathrm{GaAs}$ layer. The spin-current density $\mathbf{J}_s(x, t)$ is a second-rank tensor characterizing the direction of spin flow and the orientation and magnitude of the carried spin polarization per unit volume~\cite{Dyakonov:1971}. In the vicinity of the $\mathrm{Ni}|\mathrm{GaAs}$ interface, the density $\mathbf{J}_\mathrm{SP}(x = t_\mathrm{F}, t)$ of the spin current pumped into $\mathrm{GaAs}$ can be evaluated via the approximate relation $\mathbf{e}_n \cdot \mathbf{J}_\mathrm{SP} \simeq (\hbar / 4 \pi) \mathrm{Re} \big[ g^r_{\uparrow \downarrow} \big] \mathbf{m} \times \dot{\mathbf{m}}$, where $\mathbf{e}_n$ is the unit vector normal to the interface and pointing into $\mathrm{GaAs}$, $\hbar$ is the reduced Planck constant, $g^r_{\uparrow \downarrow}$ is the reflection spin-mixing conductance per unit area, and a small contribution caused by the imaginary part of $g^r_{\uparrow \downarrow}$ is neglected~\cite{Zwierzycki:2005}. Since $\mathrm{Re} \big[ g^r_{\uparrow \downarrow} \big]$ may be set equal to $1.5 \times 10^{17}$~m$^{-2}$ for the $\mathrm{Ni}|\mathrm{GaAs}$ interface~\cite{Ando:2011}, the above relation and the simulation data on the temporal variation of $\mathbf{m}(x = t_\mathrm{F}, t)$ enable us to evaluate the spin-current density $\mathbf{J}_\mathrm{SP}(x = t_\mathrm{F}, t)$. 

\begin{figure}
  \begin{minipage}{0.7\linewidth}
    \center{
    \includegraphics[width=1\linewidth]{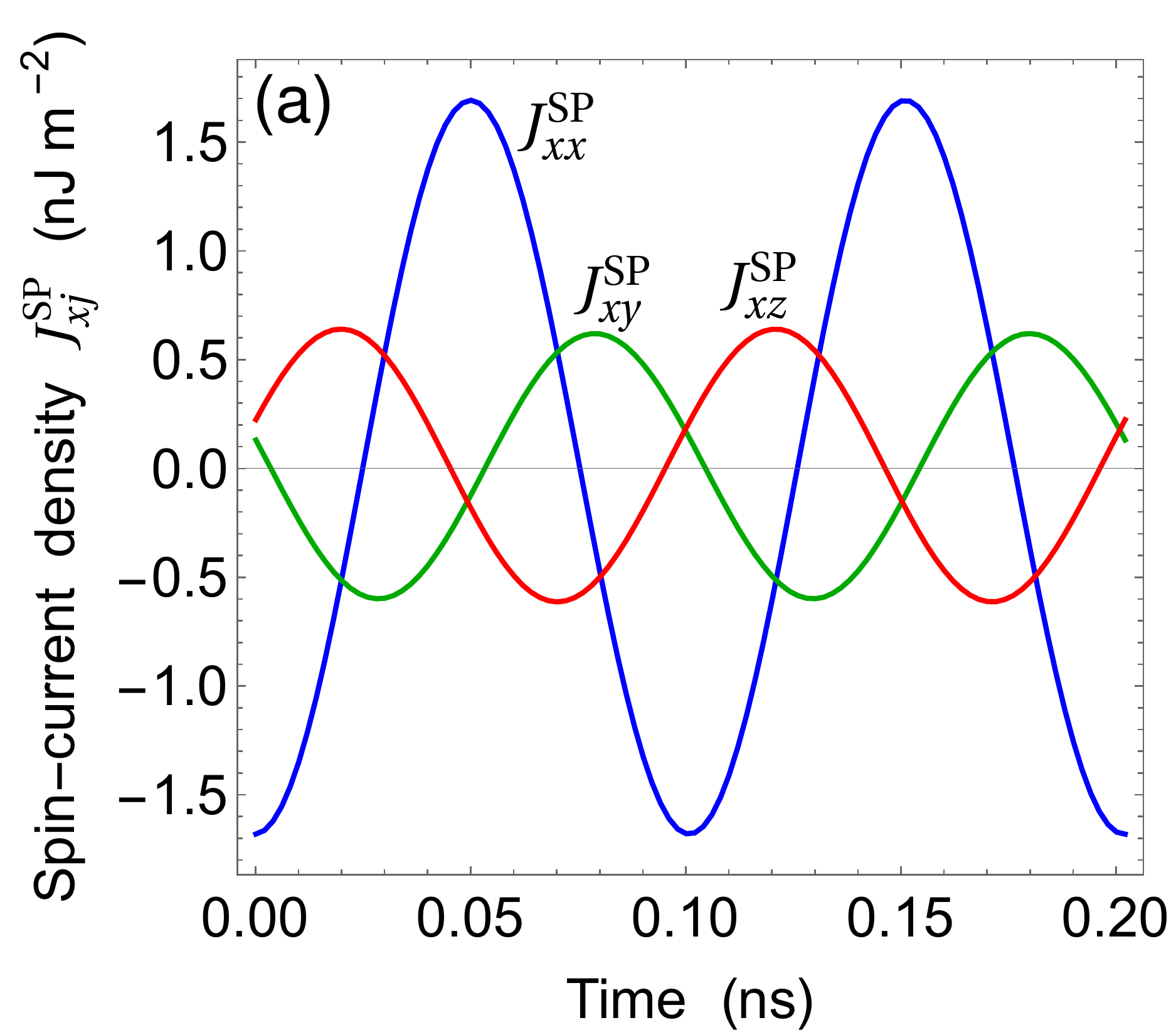}
    }
  \end{minipage}
  \begin{minipage}{0.7\linewidth}
    \center{
    \includegraphics[width=1\linewidth]{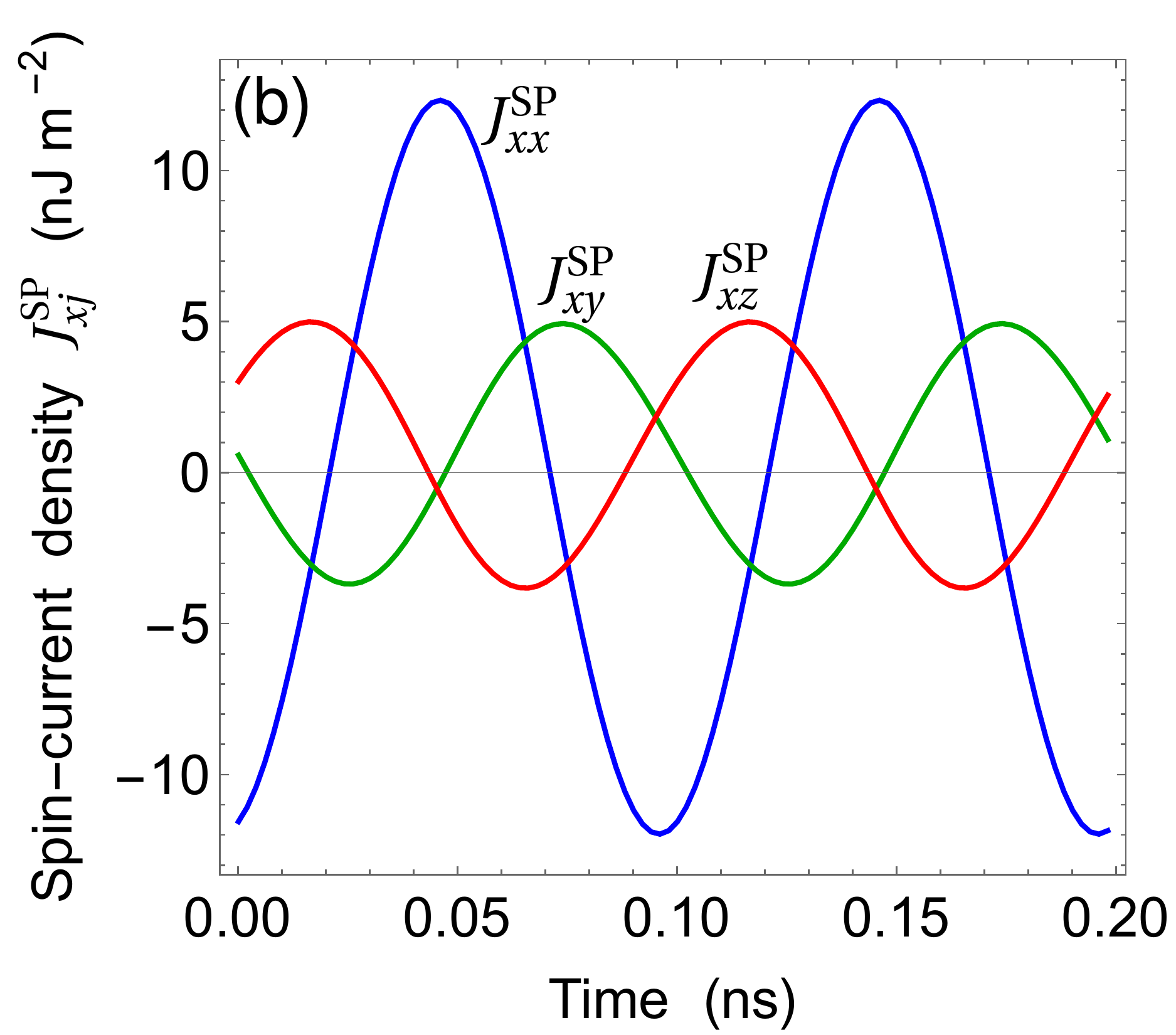}
    }
  \end{minipage}
    \caption{\label{fig:sp_vs_time}
  Time dependences of the spin-current densities $J_{xj}^\mathrm{SP}$ pumped into $\mathrm{GaAs}$ by the longitudinal (a) and transverse (b) elastic waves with the frequencies 9.9~GHz and 10~GHz, respectively. The time period shown in the figure corresponds to the steady-state regime of the magnetization precession at the $\mathrm{Ni}|\mathrm{GaAs}$ interface. The thickness of the $\mathrm{Ni}$ layer equals three quarters of the wavelength of the driving longitudinal or transverse elastic wave.}
\end{figure}

Figure~\ref{fig:sp_vs_time} shows time dependences of three nonzero components $J^\mathrm{SP}_{xj}$ of the tensor $\mathbf{J}_\mathrm{SP}$, which settle in the steady-state regime of the magnetization precession at  the excitation frequency $\nu = \nu_\mathrm{max}$. It can be seen that the transverse wave creates much stronger spin pumping into $\mathrm{GaAs}$ than the longitudinal one. For both types of elastic excitations, the amplitude of $J^\mathrm{SP}_{xx}$ is about two times larger than almost equal amplitudes of $J^\mathrm{SP}_{xy}$ and $J^\mathrm{SP}_{xz}$.
\noindent
The averaging over the period $1/\nu$ of (almost sinusoidal) spin-current variations shows that $\langle J^\mathrm{SP}_{xx} \rangle_t$ is negligible, whereas there are very small nonzero dc components $\langle J^\mathrm{SP}_{xy} \rangle_t = \langle J^\mathrm{SP}_{xz} \rangle_t$ of the pumped spin current. It should be noted that, owing to relatively small reflection spin mixing conductance of the $\mathrm{Ni}|\mathrm{GaAs}$ interface, the spin pumping into $\mathrm{GaAs}$ does not significantly increase the effective damping of the magnetization precession in $\mathrm{Ni}$~\cite{PRAppl:2020}.

The pumped spin current generates non-equilibrium spin accumulation $\bm{\mu}_s(x, t)$, which gives rise to a spin backflow at the interface with the density $\mathbf{J}_\mathrm{SB}(x = t_\mathrm{F}, t)$ amounting to $\mathbf{e}_n \cdot \mathbf{J}_\mathrm{SB} \approx - \mathrm{Re}\big[ g^r_{\uparrow \downarrow} \big] \bm{\mu}_s / 4 \pi$~\cite{Tserkovnyak:2005}. The overall spin-current density $\mathbf{J}_s(x, t)$ decays inside the $\mathrm{GaAs}$ layer due to spin relaxation and diffusion. The spatial distribution of the density $\mathbf{J}_s$ depends on that of the spin accumulation $\bm{\mu_s}$~\cite{Tserkovnyak:2002}, being defined in our one-dimensional model by the relation $\mathbf{e}_n \cdot \mathbf{J}_s(x, t) = -[\sigma \hbar / (4 e^2)] \partial \bm{\mu}_s(x, t) / \partial x$, where $e$ is the elementary positive charge, and $\sigma$ is the electrical conductivity, which amounts to $3.68 \times 10^4$~S~m$^{-1}$ for n$^+$-$\mathrm{GaAs}$~\cite{Kikkawa:1998, PRAppl:2020}. We find the spin accumulation $\bm{\mu}_s(x, t)$ by solving the diffusion equation~\cite{Tserkovnyak:2002} appended by the boundary conditions for the spin currents at the $\mathrm{Ni}|\mathrm{GaAs}$ interface $x = t_\mathrm{F}$ and the $\mathrm{GaAs}$ free surface $x = t_\mathrm{F} + t_\mathrm{N}$, which read $\mathbf{J}_s(x = t_\mathrm{F}) = \mathbf{J}_\mathrm{SP}(x = t_\mathrm{F}) + \mathbf{J}_\mathrm{SB}(x = t_\mathrm{F})$ and $\mathbf{J}_s(x = t_\mathrm{F} + t_\mathrm{N}) = 0$. The calculation yields
\begin{equation}
    \bm{\mu}_s^\omega = \frac{4 \pi e^2 \cosh{[\kappa (t_\mathrm{N} + t_\mathrm{F} - x)]}}{e^2 \mathrm{Re}\big[ g^r_{\uparrow \downarrow} \big] \cosh{(\kappa t_\mathrm{N})} + \pi \sigma \hbar \kappa \sinh{(\kappa t_\mathrm{N})}}\mathbf{e}_n \cdot \mathbf{J}_\mathrm{SP}^\omega,
    \label{eq:mu:harmonics}
\end{equation}
\noindent
where $\bm{\mu}_s^\omega$ and $\mathbf{J}_\mathrm{SP}^\omega$ denote the complex amplitudes of the harmonics having the angular frequency $\omega$, which represent the Fourier components of the spin accumulation $\bm{\mu}_s(x, t)$ and spin pumping density $\mathbf{J}_\mathrm{SP}(x = t_\mathrm{F}, t)$, and the parameter $\kappa = \lambda_\mathrm{sd}^{-1} \sqrt{1+i \omega \tau_\mathrm{sf}}$ depends on spin-diffusion length $\lambda_\mathrm{sd}$ and spin-flip relaxation time $\tau_\mathrm{sf}$. Equation~(\ref{eq:mu:harmonics}) differs from a similar relation derived in Ref.~\cite{Tserkovnyak:2002} by the account of the spin backflow. In the case of $\mathrm{GaAs}$, the spin backflow cannot be neglected because it appears to be rather strong at $\lambda_\mathrm{sd} = 2.32$~$\mu$m~\cite{Kikkawa:1998} and $\tau_\mathrm{sf} = 0.9$~ns~\cite{Bhat:2014}. Since the elastically driven spin pumping is almost monochromatic in our setting, Eq.~(\ref{eq:mu:harmonics}) is valid for the sought relation between $\bm{\mu}_s(x, t)$ and $\mathbf{J}_\mathrm{SP}(x = t_\mathrm{F}, t)$ as well, which enables us to calculate the overall spin-current density $\mathbf{J}_s(x, t)$.

Owing to the inverse spin Hall effect (ISHE), the spin current in the $\mathrm{GaAs}$ layer generates a charge current with the density $\mathbf{J}_c^\mathrm{ISHE}$ defined by the formula~\cite{Mosendz:2010}
\begin{equation}
    \mathbf{J}_c^\mathrm{ISHE}(x, t) = \alpha_\mathrm{SH} (2e / \hbar) \mathbf{e}_n \times [\mathbf{e}_n \cdot \mathbf{J}_s(x, t)],
    \label{eq:Jc}
\end{equation}
\noindent
where $\alpha_\mathrm{SH} = 0.007$ is the spin Hall angle of $\mathrm{GaAs}$~\cite{Ando:2011}. Under considered open-circuit electrical boundary conditions, the transverse charge current $\mathbf{J}_c^\mathrm{ISHE}$ flowing along the interface should create a charge accumulation at the lateral boundaries of the $\mathrm{GaAs}$ film. Such an accumulation induces an electric field $\mathbf{E}$ in $\mathrm{GaAs}$, which causes a drift current with the density $\mathbf{J}_c^\mathrm{drift} = \sigma \mathbf{E}$. To calculate the spatial distribution of the electric potential $\varphi$ in the $\mathrm{Ni}$/$\mathrm{GaAs}$ bilayer and the total charge current density $\mathbf{J}_c = \mathbf{J}_c^\mathrm{ISHE} + \mathbf{J}_c^\mathrm{drift}$, we numerically solve the Laplace's equation $\nabla^2 \varphi = 0$ with the appropriate boundary conditions. The latter follow from the absence of charge current across the outer surfaces of the bilayer, and the absence of $\mathbf{J}_c^\mathrm{ISHE}$ inside $\mathrm{Ni}$. It should be noted that the potential $\varphi$ should be regarded as a complex quantity since the parameter $\kappa$ affecting the spin-current density $\mathbf{J}_s$ involved in Eq.~(\ref{eq:Jc}) has a substantial complex part at $\omega \tau_\mathrm{sf} >> 1$.

\begin{figure}[h]
  \begin{minipage}{0.7\linewidth}
    \center{
    \includegraphics[width=0.9\linewidth]{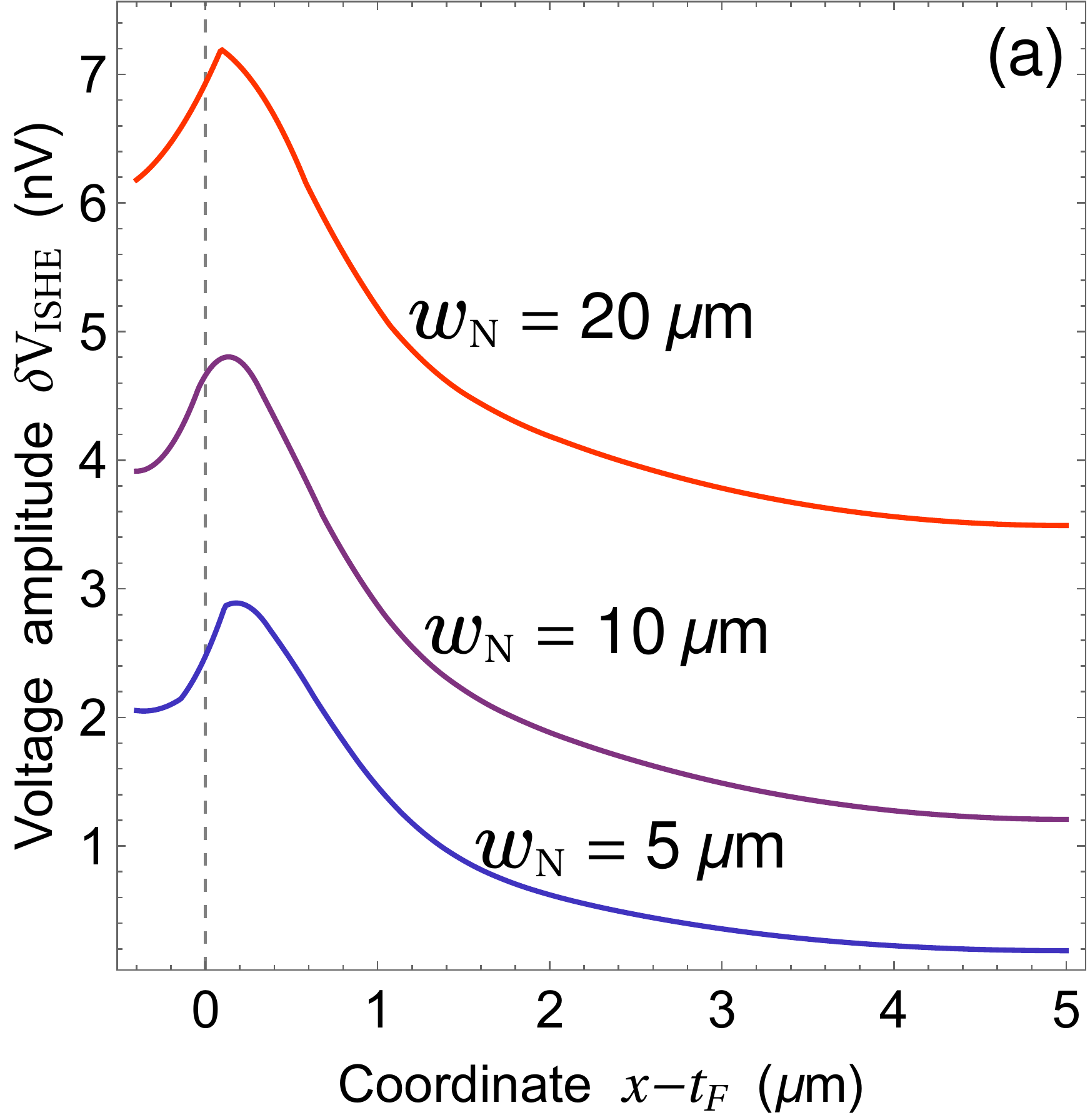}
    }
  \end{minipage}
  \begin{minipage}{0.7\linewidth}
    \center{
    \includegraphics[width=0.9\linewidth]{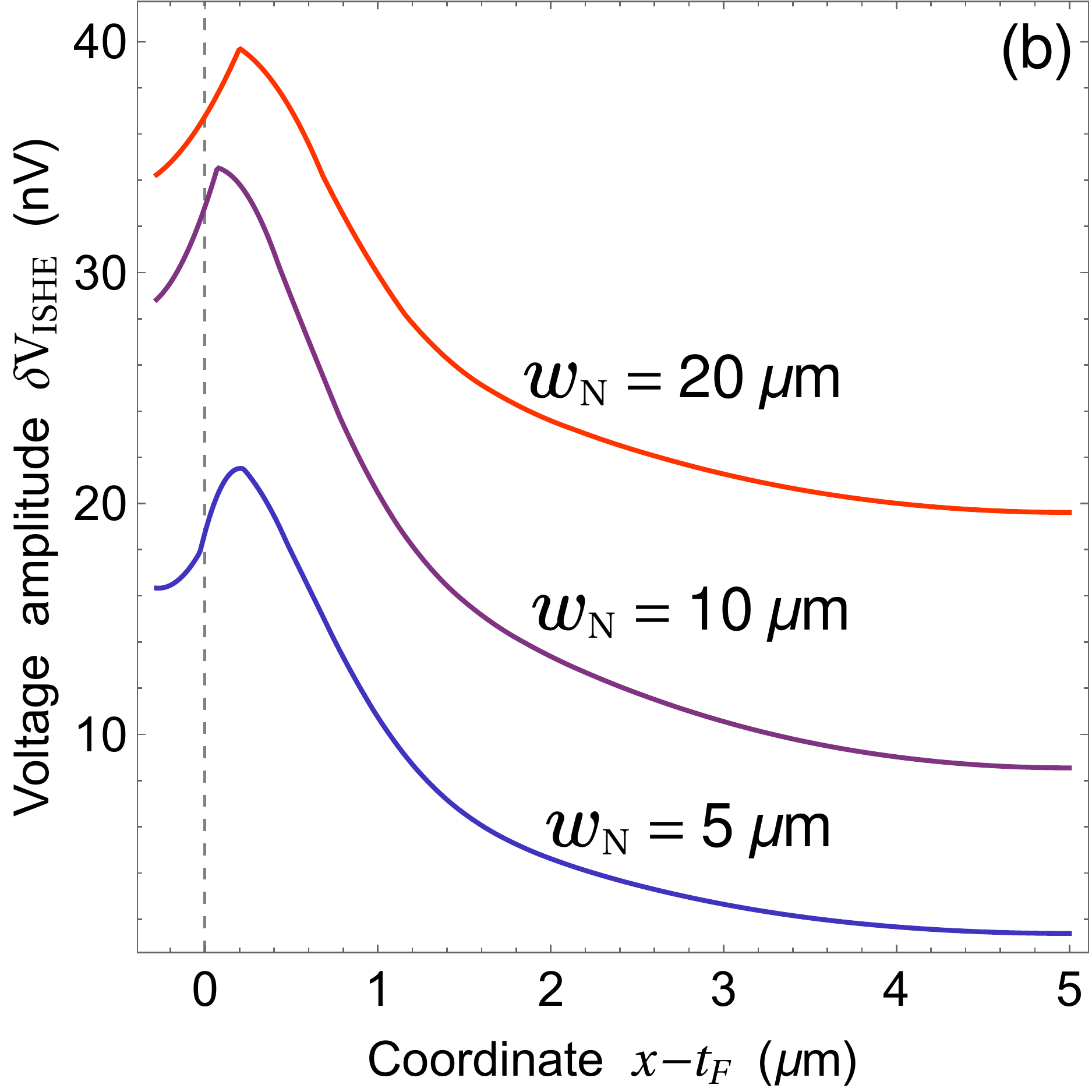}
    }
  \end{minipage}
    \caption{\label{fig:Vishe}
  Amplitude $\delta V_z^\mathrm{ISHE}(x)$ of the ac voltage between the lateral sides of the $\mathrm{Ni}$/$\mathrm{GaAs}$ bilayer excited by longitudinal (a) and transverse (b) elastic waves with the frequencies 9.9~GHz and 10~GHz, respectively. The thickness of the $\mathrm{GaAs}$ layer equals 5~$\mu$m, and its width $w_\mathrm{N}$ is indicated in the figure. }
\end{figure}

In the numerical calculations, we consider only the component $J^s_{xy}$ of the elastically generated spin current $\mathbf{J}_s$, because $J^s_{xx}$ does not create any charge flow, and the components $J^s_{xy}$ and $J^s_{xz}$ have almost equal magnitudes and can be probed independently via transverse voltages $V_z^\mathrm{ISHE} = \varphi(z = w_\mathrm{N}/2) - \varphi(z = -w_\mathrm{N}/2)$ and $V_y^\mathrm{ISHE} = \varphi(y = 0) - \varphi(y = -h_\mathrm{N})$, respectively (Fig.~\ref{fig:setup}). Figure~\ref{fig:Vishe} shows the amplitude $\delta V_z^\mathrm{ISHE}(x)$ of the oscillating voltage $V_z^\mathrm{ISHE}(x, t)$ calculated at the excitation frequency $\nu = \nu_\mathrm{max}$ for the $\mathrm{GaAs}$ films with the thickness $t_\mathrm{N} = 5$~$\mu$m.

It can be seen that $\delta V_z^\mathrm{ISHE}$ varies nonmonotonically with the distance $x-t_\mathrm{F}$ from the $\mathrm{Ni}|\mathrm{GaAs}$ interface, reaching its maximum inside the semiconductor at $x-t_\mathrm{F} = 100-200$~nm. The voltage amplitude $\delta V_z^\mathrm{ISHE}(x)$ grows with the increasing width $w_\mathrm{N}$ of the $\mathrm{GaAs}$ film, and the voltage peak becomes much higher at the excitation of magnetization dynamics by the shear elastic wave (see Fig.~\ref{fig:Vishe}). Importantly, the transverse ac voltage $V_z^\mathrm{ISHE}$ characterizing the spin pumping induced by either type of elastic waves is high enough for the experimental measurement near the $\mathrm{Ni}|\mathrm{GaAs}$ interface.

\begin{figure}
    \center{
        \includegraphics[width=0.8\linewidth]{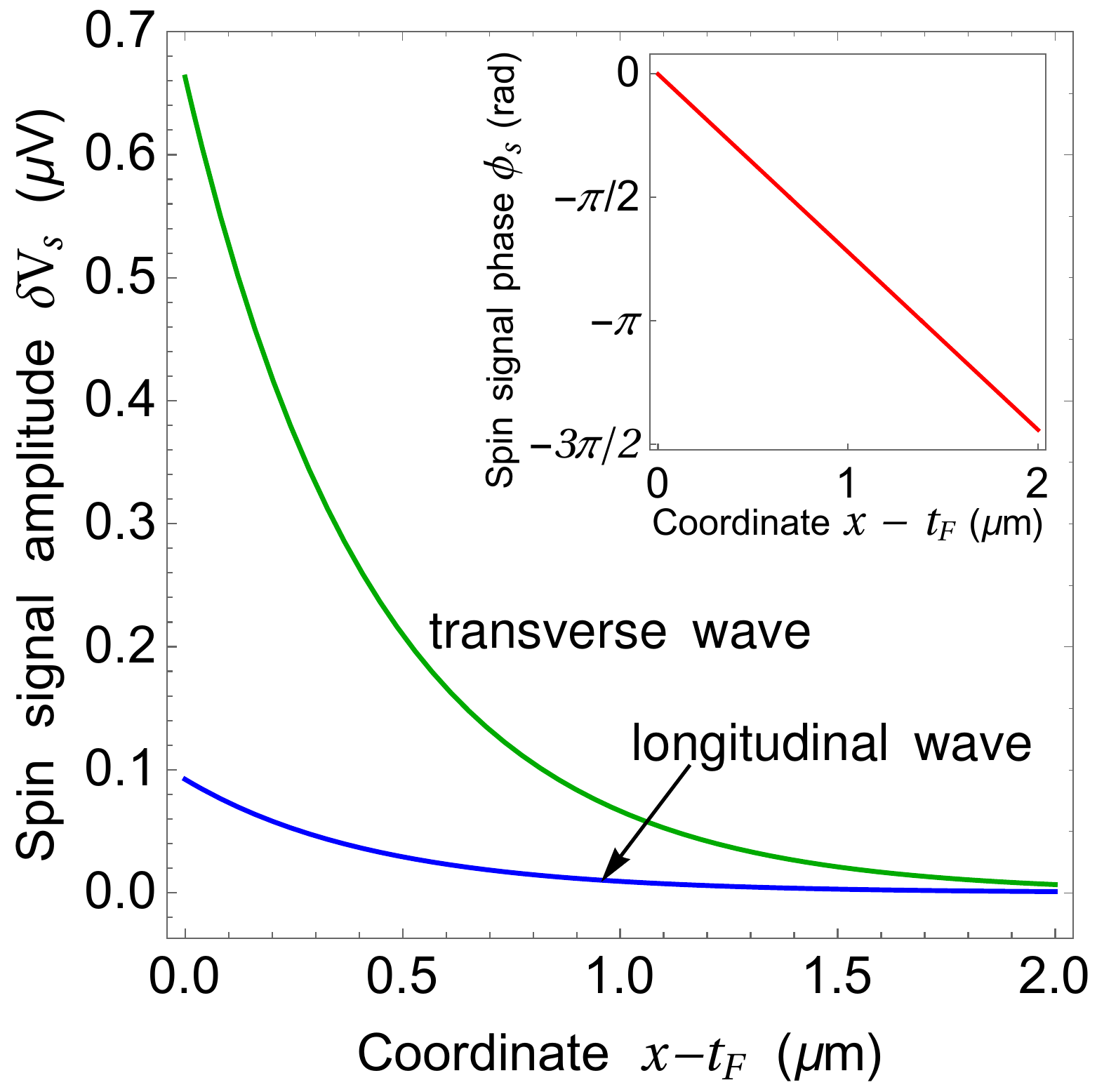}
    }
    \caption{\label{fig:Vs}
  Amplitude $\delta V_s$ and phase $\phi_s$ (inset) of the ac spin voltage between the lateral Fe probe and a normal-metal contact at the $\mathrm{GaAs}$ free surface plotted as a function of the distance $x-t_\mathrm{F}$ from the $\mathrm{Ni}|\mathrm{GaAs}$ interface. The $\mathrm{Ni}$/$\mathrm{GaAs}$ bilayer is excited either by the longitudinal wave with frequency 9.9~GHz or by the transverse wave with frequency 10~GHz. The thickness of the $\mathrm{GaAs}$ layer equals 5~$\mu$m. }
\end{figure}

Another method to evaluate the spin pumping into a normal metal or semiconductor experimentally is known as a nonlocal spin detection scheme~\cite{Silsbee:1985, Lou:2007}. This scheme measures a voltage $V_s$ between a ferromagnetic probe and a nonmagnetic electrode brought into contact with the semiconductor. Since the voltage $V_s$ is directly proportional to the product $\bm{\mu}_s \cdot \mathbf{M}_\mathrm{probe}$, where $\mathbf{M}_\mathrm{probe}$ is the probe magnetization, it is possible to detect all three components of the vector $\bm{\mu}_s$ by using differently magnetized ferromagnetic contacts. As a representative example, we consider an iron probe magnetized along the $x$ axis, which is placed on the lateral side of the $\mathrm{GaAs}$ layer (Fig.~\ref{fig:setup}), and a normal-metal electrode deposited on the free surface $x = t_\mathrm{F} + t_\mathrm{N}$ of the 5-$\mu$m-thick $\mathrm{GaAs}$ film. In this case, the spin voltage $V_s(x, t)$ is defined by the relation $V_s(x, t) = \eta_\mathrm{IE} p_\mathrm{Fe} \mu_x^s(x, t) / (2e)$, where $\eta_\mathrm{IE}$ is the spin transmission efficiency of the $\mathrm{GaAs}|\mathrm{Fe}$ interface, $p_\mathrm{Fe}$ is the spin polarization of $\mathrm{Fe}$ at the Fermi level, and the spin accumulation $\mu_x^s$ beneath the probe with nanoscale dimensions is assumed uniform. Figure~\ref{fig:Vs} shows the amplitude $\delta V_s(x)$ and phase $\phi_s(x)$ of the ac spin voltage calculated using the parameters $\eta_\mathrm{IE} \approx 0.5$ and $p_\mathrm{Fe} \approx 0.42$ characteristic of the Schottky tunnel barrier between Fe probe and n$^+$-$\mathrm{GaAs}$~\cite{Lou:2007}.
Importantly, the voltage amplitude $\delta V_s$ appears to be rather large near the $\mathrm{Ni}|\mathrm{GaAs}$ interface, exceeding 650~nV under the excitation by the transverse elastic wave and 80~nV in the bilayer excited by the longitudinal wave (see Fig.~\ref{fig:Vs}). Although $\delta V_s(x)$ gradually decreases with the increasing distance $x-t_\mathrm{F}$ from the $\mathrm{Ni}|\mathrm{GaAs}$ interface, it remains to be measurable experimentally even at the distances over 0.5~$\mu$m. The phase $\phi_s(x)$ of the spin voltage varies linearly inside the $\mathrm{GaAs}$ layer and changes strongly already at $x-t_\mathrm{F} \sim 0.5$~$\mu$m (Fig.~\ref{fig:Vs}). This result demonstrates that the phase difference between the spin accumulation inside $\mathrm{GaAs}$ and the spin pumping at the $\mathrm{Ni}|\mathrm{GaAs}$ interface may be large owing to the condition $\omega \tau_\mathrm{sf} >> 1$.

The input electric power $W$ necessary for the functioning of the proposed spin injector can be estimated from the generated acoustic power using the relation $W = \frac{1}{K^2}\frac{1}{2} A \rho c \omega^2 u_{\text{max}}^2$, where $K^2$ is the electromechanical transduction efficiency of the piezoelectric transducer \cite{Bhaskar:2020}, $c$ is the velocity of the generated longitudinal or transverse acoustic wave, and $A$ denotes the dynamically strained area of the ferromagnetic film having the mass density $\rho$. Expressing $u_{\text{max}}$ via the maximal strain $\varepsilon_{\text{max}}$ in the acoustic wave, we obtain $W = \alpha \frac{1}{K^2}\frac{1}{2} A\rho c^3 \varepsilon_{\text{max}}^2$. For the device with $A < 25~\mu$m$^2$ and $K^2 = 12\%$ \cite{Bhaskar:2020}, which is driven by a transverse ($\alpha=4$) or longitudinal ($\alpha=1$) wave with $\varepsilon_{\text{max}} = 10^{-4}$, the calculation yields $W < 2$ mW. This value is much smaller than the lowest power consumption $W \approx 25$ mW of the spin injector driven by the microwave magnetic field \cite{Ando:2011}.

\section{Conclusions}
In this work, we theoretically studied the coupled elastic and spin dynamics induced in $\mathrm{Ni}$/$\mathrm{GaAs}$ bilayers by longitudinal and transverse acoustic waves generated by the attached piezoelectric transducer (Fig.~\ref{fig:setup}). Using advanced micromagnetoelastic simulations, we first modeled the elastically driven magnetization dynamics in thick $\mathrm{Ni}$ films at the wave frequencies around the resonance frequency $\nu_\mathrm{res}$ of the coherent magnetization precession in unstrained $\mathrm{Ni}$ film. The simulations showed that this dynamics has the form of a forced spin wave having the frequency and wavelength of the monochromatic driving wave. Remarkably, the transverse elastic wave creates much stronger spin wave than the longitudinal one at the considered external magnetic field (Fig.~\ref{fig:waves_all}). The backaction of travelling spin wave on the elastic dynamics manifests itself in the generation of weak secondary elastic waves created by the magnetization precession (see Fig.~\ref{fig:waves_secondary}). These waves are characterized by oscillating strains $\varepsilon_{ij}(x, t)$ different from the strain $\varepsilon_{xx}(x, t)$ or $\varepsilon_{xz}(x, t)$ in the primary driving wave, and they were not reported in the previous work on the modeling of magnetization dynamics induced by longitudinal elastic waves in $\mathrm{Ni}$~\cite{Chen:2017}. The magnetoelastic feedback also influences the driving elastic wave, leading to a gradual reduction of its amplitude during the propagation in $\mathrm{Ni}$, which adds to the ``acoustic'' decay caused by the wave attenuation of electronic origin~\cite{Homer:1987}. At the considered wave frequencies $\nu \approx 10$~GHz, the decay resulting from the energy transfer to the magnetic subsystem is stronger than the acoustic decay for the transverse waves but small for longitudinal ones. Importantly, both types of elastic waves are expected to carry spin signals over significant distances of several micrometers in $\mathrm{Ni}$.

We also modeled the magnetoelastic dynamics of $\mathrm{Ni}$/$\mathrm{GaAs}$ bilayers at the excitation frequencies $\nu \sim \nu_\mathrm{res}$, focusing on the $\mathrm{Ni}$ thicknesses comparable to the wavelength of the injected acoustic wave. The simulations allowed for the reflections of the elastic waves from the boundaries of the $\mathrm{Ni}$ layer and demonstrated the excitation of a nonhomogeneous magnetization dynamics in it. Importantly, a steady-state magnetization precession with frequency equal to the excitation frequency and constant amplitude was revealed at the $\mathrm{Ni}|\mathrm{GaAs}$ interface after a short transition period of about 1~ns (Fig.~\ref{fig:precession_interface}). The simulations performed for $\mathrm{Ni}$ layers of different thicknesses showed that the amplitude of stationary precession has a maximum at $\mathrm{Ni}$ thickness amounting to three quarters of the driving elastic wave wavelength. This finding, which differs from the results of simulations carried out for $\mathrm{Fe}_{81}\mathrm{Ga}_{19}$/$\mathrm{Au}$ and $\mathrm{CoFe}_2\mathrm{O}_4$/$\mathrm{Pt}$ bilayers~\cite{APL:2017, PRB:2019}, was explained by an analytical model giving simple criteria for the optimal geometry of an elastic bilayer that maximizes the strain amplitude at the interface.

Numerical results obtained for the steady-state magnetization precession at the $\mathrm{Ni}|\mathrm{GaAs}$ interface were used to evaluate the spin-current densities pumped into $\mathrm{GaAs}$ by the dynamically strained $\mathrm{Ni}$ film (Fig.~\ref{fig:sp_vs_time}). The spin accumulation in the semiconductor was then calculated by solving numerically the spin diffusion equation with the account of the spin pumping into $\mathrm{GaAs}$ and the spin backflow into $\mathrm{Ni}$. Since the spin current creates a charge current owing to the ISHE, the spin generation in $\mathrm{GaAs}$ can be detected via electrical measurements. Therefore, we also determined the distribution of the electric potential in the $\mathrm{Ni}$/$\mathrm{GaAs}$ bilayer with open-circuit electrical boundary conditions by numerically solving the Laplace's equation. This enabled us to evaluate the transverse voltage appearing between the lateral sides of the dynamically strained $\mathrm{Ni}$/$\mathrm{GaAs}$ bilayer. It was shown that the amplitude of this ac voltage is large enough for the experimental detection near the $\mathrm{Ni}|\mathrm{GaAs}$ interface (Fig.~\ref{fig:Vishe}). Furthermore, spin accumulation manifests itself in the voltage between a ferromagnetic probe and a nonmagnetic electrode brought into contact with the semiconductor (Fig.~\ref{fig:setup}). Performing calculations of this ac spin voltage, we found that it retains measurable amplitude even at the distances over 0.5~$\mu$m from the $\mathrm{Ni}|\mathrm{GaAs}$ interface (Fig.~\ref{fig:Vs}).

Thus, our theoretical study of the $\mathrm{Ni}$/$\mathrm{GaAs}$ heterostructure demonstrated that the spin injector employing elastic waves is promising for the spin generation in semiconductors. Since the proposed device can be driven electrically via the strain-mediated magnetoelectric effect, it has much lower power consumption than the spin injector excited by a microwave magnetic field~\cite{Ando:2011}.

\section{Acknowledgements} 
The work was supported by the Foundation for the Advancement of Theoretical Physics and Mathematics "BASIS".

\bibliography{References}
\end{document}